\documentclass[12pt]{article}

\usepackage{graphicx}
\begin{document}

\baselineskip=16.4pt plus 0.2pt minus 0.1pt

\makeatletter
\@addtoreset{equation}{section}
\renewcommand{\theequation}{\thesection.\arabic{equation}}
\newcommand{\abs}[1]{\left| #1 \right|}
\newcommand{\frbox}[2]{ \fbox{\parbox{ #1 }{ #2 }} }
\newcommand{\nn}{\nonumber\\}
\newcommand{\ket}[1]{\left| #1 \right>}
\newcommand{\kakko}[1]{\left( #1 \right)}
\newcommand{\ckakko}[1]{\left\{ #1 \right\}}
\newcommand{\dkakko}[1]{\left[ #1 \right]}
\newcommand{\dz}[1]{\frac{d #1}{2\pi i}}
\newcommand{\fron}[1]{\frac{1}{#1}}
\newcommand{\Q}{Q_{\rm B}}
\newcommand{\rL}{{\rm L}}
\newcommand{\rR}{{\rm R}}
\newcommand{\rC}{{\rm C}}
\newcommand{\CL}{{C_\rL}}
\newcommand{\CR}{{C_\rR}}
\newcommand{\JB}{J_{\rm B}}
\newcommand{\Jgh}{J_{gh}}
\newcommand{\reg}{{\rm reg.}}
\newcommand{\Cl}{{{\rm C}_{\rm left}}}
\newcommand{\Cr}{{{\rm C}_{\rm right}}}

\begin{titlepage}
\title{
\hfill{\normalsize hep-th/0502042}\\
\vspace{4ex}
\bf Classical Solutions and Order of Zeros\\ in Open String Field
Theory\vspace{.5cm}} 
\author{
Yuji {\sc Igarashi}$^{1}$\thanks{E-mail address:
igarashi@ed.niigata-u.ac.jp}\ ,\ \ Katsumi {\sc Itoh}$^{1}$\thanks{E-mail
address: itoh@ed.niigata-u.ac.jp}
\vspace{.2cm}\\
Fumie {\sc Katsumata}$^{2}$\thanks{E-mail
address: katsumata@asuka.phys.nara-wu.ac.jp}\ ,\ \ Tomohiko
{\sc Takahashi}$^{2}$\thanks{E-mail
address: tomo@asuka.phys.nara-wu.ac.jp}\ \ and Syoji
{\sc Zeze}$^{3}$\thanks{E-mail
address: zeze@yukawa.kyoto-u.ac.jp}
\vspace{3ex}\\
$^1${\it Faculty of Education, Niigata University, Japan}\\
$^2${\it Department of Physics, Nara Women's University, Japan}\\
$^3${\it Yukawa Institute for Theoretical Physics,
Kyoto University, Japan}}
 
\date{\normalsize January, 2005}
\maketitle
\thispagestyle{empty}
\vspace{.5cm}

\begin{abstract}
\normalsize

Earlier an analytic approach is proposed for classical solutions
describing tachyon vacuum in open string field theory.
Based on the approach, we construct a certain class
of classical solutions written in terms of holomorphic functions with
higher order zeros.  Taking the simplest among the new classical
solutions, we study the cohomology of the new BRS charge and make a
numerical analysis of the vacuum energy.  The results indicate that the
new non-trivial solution is another analytic candidate for the tachyon
vacuum.

\end{abstract}
\end{titlepage}

\section{Introduction}

Over the past years, it has been argued that string field theory,
originally defined around the D-brane vacuum, admits the lower energy
vacuum where the D-brane is annihilated via the condensation of the open
string tachyon \cite{rf:Sen-review,rf:Ohmori}. This should be described
by a classical solution for which the value of the action is the
minus of the tension of the annihilated brane \cite{rf:Sen,rf:SenUniv}.
The classical solution in cubic open string field theory (CSFT)
\cite{rf:CSFT} has been constructed numerically using the level
truncation method in the Siegel gauge
\cite{rf:KS-tachyon,rf:SZ-tachyon,rf:MT,rf:GR}.  These studies provide
us with sufficiently accurate quantitative tests of the above
conjecture.

Remarkably, there exists a candidate for analytic description of the
tachyon condensed vacuum as classical solutions in CSFT.  In
\cite{rf:TT}, one parameter family of classical solutions in pure gauge
form has been constructed in a half string formulation by taking certain
linear combinations of the BRS current and the ghost field acting on the
identity operator.  A linear combination is specified by a holomorphic
function $h_{a}(w)$ with a real parameter $a$.  The solutions are
expressed on the universal basis: the matter Virasoro operators, the
ghost and anti-ghost fields.  The work \cite{rf:TT} is extended in
\cite{rf:KT} to give more general solutions by considering infinite set
of one parameter families of holomorphic functions $h_{a}^{m}(w)$
labelled by a positive integer $m$.  The solutions are in the pure gauge
form and mostly reduced to gauge transformations of the usual vacuum.
However non-trivial solutions emerge when the parameter in
$h_{a}^{m}(w)$ takes the boundary value $a=a_{b}$.  Thus, we have a
series of non-trivial classical solutions labelled by the integer $m$.

Given a solution with $h_{a}^{m}(w)$, we may expand the action around
the solution to have a theory for the string fluctuating around the
classical solution.  In CSFT, the new action remains the same form, but
the BRS charge $Q_{B}$ is replaced by the new one ($Q'_{B}$), which
depends on the function $h_{a}^{m}$. Study on the cohomology defined by
$Q'_{B}$ \cite{rf:KT}, perturbative scattering amplitudes \cite{rf:TZ1}
and a numerical analysis of the vacuum energy \cite{rf:tomo} strongly
indicate that the non-trivial solutions correspond to the tachyon
vacuum.

It is found that properties of the analytic solutions depend on the
structure of zeros in the underlying function $F(w) = e^{h(w)}$
\cite{rf:Drkr2,rf:zeze}: the functions $F_{a}^{m}(w)$ for $a \neq
a_{b}$ have $4m$ first-order zeros inside as well as outside of the unit
circle $|w| =1$.  In the limit $a \to a_{b}$, each pair of zeros, one
inside and the other outside, becomes degenerate on the unit circle. The
functions $F_{a_{b}}^{m}(w)$ turn out to have $2m$ second-order zeros
on the unit circle, and produce non-trivial solutions which cannot be
gauged away.  The implications of the structure of zeros in the function
$F(w)$ remain to be clarified. One asks \\ (1) how the solutions with
$F_{a_{b}}^{l}$ and $F_{a_{b}}^{m}~~(l \neq m)$, which differ in
locations and numbers of zeros, could be related,\\ (2) if some new aspects
appear in another class of solutions associated with functions having
higher order zeros, in particular, if those solutions give rise to
different values for the vacuum energy, reflecting the order of zeros.

The purpose of this paper is to consider the above question (2). We
construct a classical non-trivial solution to CSFT characterized by a
holomorphic function with fourth order zeros, and discuss its properties
in details.  After obtaining explicit form of the solution, we examine
cohomology of the new BRS charge, the kinetic operator for shifted
string field around the solution. Numerical analysis of the vacuum
energy is also reported. Our results indicate that the new non-trivial
solution with higher order zeros provides us with yet another analytic
candidate for the tachyon vacuum.

\section{Classical Solutions in Open String Field Theory}

Let us briefly summarize the approach in Ref. \cite{rf:TT} and its
results on classical solutions in CSFT.  The action of CSFT
\cite{rf:CSFT} is given by
\begin{eqnarray}
 S[\Psi]=-\frac{1}{g^2}\int \left(\frac{1}{2}\Psi*\Q\Psi
+\frac{1}{3}\Psi*\Psi*\Psi\right),
\label{Eq:action}
\end{eqnarray}
where $\Psi$ is the string field and $\Q$ is the Kato-Ogawa 
BRS charge \cite{rf:KO}. This action leads to the equation of motion, 
\begin{eqnarray}
 \Q\Psi+\Psi*\Psi=0.
\label{Eq:eqmotion}
\end{eqnarray}
The classical solutions are written in terms of two basic operators, the
BRS current $J_{\rm B}(w)$ and the ghost field $c(w)$, acting on the
identity operator $I$.  A solution is given by a contour integral of a
linear combination of these operators specified by a holomorphic
function $h(w) = \log F(w)$ \cite{rf:TT}:
\begin{eqnarray}
\Psi_0(h)&=& \int_{C_{\rm left}}\frac{dw}{2\pi i}~\left[
\left(e^{h}-1 \right)~J_{\rm B}(w)
- \left((\partial h)^{2} e^{h}  \right)~c(w) \right] I\nonumber\\ 
&\equiv& Q_{\rm L}(e^{h}-1)I-C_{\rm L}\left((\partial h)^{2} e^{h} \right)I.
\label{Eq:sol}
\end{eqnarray}
Here, the integral is taken along the left-half of a circle, as
indicated by $C_{\rm left}$ \cite{rf:TT}.  The function $h(w)$ is
required to satisfy two conditions, (i) $h(-1/w) = h(w)$ and (ii) $h(\pm
i) =0$.  The first condition is necessary to ensure the basic symmetry
property of CSFT under the inversion and the second one guarantees the
closure of the half-splitting algebra.

Expanding the original string field around the classical solution,
\begin{eqnarray}
 \Psi= \Psi_{0}(h) +\Phi,
\end{eqnarray}
we rewrite the action as
\begin{eqnarray}
 S[\Psi]=S[\Psi_{0}(h)]
-\frac{1}{g^2}\int \left(\frac{1}{2}\Phi*\Q' \Phi
+\frac{1}{3}\Phi*\Phi*\Phi\right).
\end{eqnarray}
Note here that the three-string vertex for the string field $\Phi$
remains the same form as the original one, while the kinetic term
carries the new BRS charge given by
\begin{eqnarray}
 \Q' \Phi&\equiv& \Q \Phi +\Psi_{0}(h)* \Phi+ \Phi*\Psi_{0}(h).
\label{Eq:newBRS1}
\end{eqnarray}
More explicitly, it is written as
\begin{eqnarray}
 \Q' &=& \oint\frac{dw}{2\pi i}~\left[e^{h}\,J_{\rm B}(w)  
- (\partial h)^{2} e^{h}\,c(w) \right] \nonumber\\
&\equiv& Q(e^{h})-C\left((\partial h)^{2} e^{h} \right),
\end{eqnarray}
where the integration is taken over a closed circle or $C_{\rm
left}+C_{\rm right}$: this is due to the contributions from the second
and third terms of eq.~(\ref{Eq:newBRS1}), in which the classical
solution is multiplied on the left and right halves of the string field
$\Psi$.

In \cite{rf:KT}, the universal solutions constructed with a set of functions 
\begin{eqnarray}
h_{a}^{m}(w)&=&\log\left\{
1-\frac{a}{2}(-1)^m\left(
w^m-\left(-\frac{1}{w}\right)^m\,\right)^2\right\}\ \ \ (m=1,2,3,\cdots)
\label{function-h}
\end{eqnarray}
were discussed. The parameter $a$ must be $a \geq -1/2$ to satisfy the
hermiticity requirement. Exponentiating these functions, we have 
\begin{eqnarray}
F_{a}^{m}(w) &=& \exp\Bigl(h_{a}^{m}(w)\Bigr) = \frac{\Bigl(1-(-1)^m Z(a)~w^{2m}\Bigr)\Bigl(1-(-1)^m Z(a)~w^{-2m}\Bigr)}
{(1-Z(a))^2},\nonumber \\
Z(a)&\equiv&\frac{1+a-\sqrt{1+2a}}{a}.
\label{function-F}
\end{eqnarray}
For $a>-1/2$, $Z(a) \in (-1,~1)$ and $Z(-1/2)=-1$. 

For each function $h_{a}^{m}(w)$, we have shown the followings for
$a>-1/2$:
\begin{enumerate}
 \item The action obtained by expanding around the solution 
       can be transformed back to the action with the
       original BRS charge \cite{rf:TT};
 \item The new BRS charge gives rise to the cohomology
       which has one-to-one correspondence to the cohomology of the
       original BRS charge \cite{rf:KT};
 \item The expanded theory reproduces the same open string scattering
       amplitudes as the original theory \cite{rf:TZ1};
 \item We can also show numerically that the expanded theory has a
       non-perturbative vacuum and its vacuum energy tends to the value
       appropriate to cancel the D-brane tension as the truncation level
       increases \cite{rf:tomo}.
\end{enumerate}
All these facts are consistent with the expectation that solutions for
$a>-1/2$ are trivial pure gauge solutions.  On the other hand, around a
solution with $a=-1/2$, we find completely different properties in the
expanded theory:
\begin{enumerate}
 \item[5.] The new BRS charge has the vanishing cohomology in the
	   Hilbert space of the ghost number one \cite{rf:KT};
 \item[6.] Any open string scattering amplitudes vanish
	   and perturbatively there are no open string excitations ({\it
	   no open string theorem}) \cite{rf:TZ1};
 \item[7.] A numerical analysis shows that the non-perturbative vacuum
	   found for $a>-1/2$ disappears as $a$ approaches to $-1/2$
	   \cite{rf:tomo}.
\end{enumerate}
Hence, we believe that the non-trivial solutions correspond to the
tachyon vacuum.

The above results are related to the distribution and the order of zeros
in the functions $\exp (h_{a}^{m}(w)) =F_{a}^{m}(w)$.  Consider the
function $F_a^{1}(w)$, for example.  For $a>-1/2$, it has the
first-order zeros at $w=\pm \sqrt{-Z(a)},\ \pm 1/\sqrt{-Z(a)}$.  For
$a=-1/2$, $Z(-1/2)=-1$ and the zeros at $\sqrt{-Z(a)}$ and
$-\sqrt{-Z(a)}$ coincide with those at $1/\sqrt{-Z(a)}$ and
$-1/\sqrt{-Z(a)}$, respectively: In other words, these points become
the second-order zeros.  We have the non-trivial solution when this
change in the distribution of zeros occurs.  In general, the function
$F_a^{m}(w)$ has $4m$ first-order zeros for $a>-1/2$. Half of them are
distributed inside the unit circle $|w|=1$ while the other half are
located outside of it, as a result of the condition
$h_{a}^{m}(w)=h_{a}^{m}(-1/w)$.  In the limit of $a \to -1/2$, these
$4m$ first-order zeros merge into $2m$ second-order zeros on the unit
circle and non-trivial solutions emerge.

From these observations, we wonder if the order of zeros has any
relevance in characterizing non-trivial solutions.  Actually, it has been
conjectured that the order is related to the number of D-branes
\cite{rf:Drkr2}.  So it
would be worthwhile to consider other solutions written in terms of
holomorphic functions with higher oder zeros.

\section{Classical Solutions with Higher Order Zeros}

We now construct a new family of classical solutions.  The underlying
holomorphic function is expressed as
\begin{eqnarray}
 h_{\{a_m\}}(w)=\log\left[
1+\sum_m a_m \Bigl(\exp\,h_{-1/2}^m(w)-1\Bigr)\right].
\end{eqnarray}
where $h_a^m(w)$ is given by eq.~(\ref{function-h}). Hence, as required
for classical solutions, $h_{\{a_m\}}(w)$ satisfies
$h_{\{a_m\}}(-1/w)=h_{\{a_m\}}(w)$ and $h_{\{a_m\}}(\pm i)=0$. The
classical solutions associated with $h_{\{a_m\}}(w)$ are parameterized
by $\{a_1,\,a_2,\cdots\}$, which can be chosen in such a way that new
solutions with higher order zeros appear.

For example, consider the function made of $h_{-1/2}^1$ and
$h_{-1/2}^2$, 
\begin{eqnarray}
 h_{\{a_1,\,a_2\}}(w)=\log\left[
1+ a_1 \left\{
-\frac{1}{4}\left(w-\frac{1}{w}\right)^2-1\right\}
+a_2 \left\{
\frac{1}{4}\left(w^2+\frac{1}{w^2}\right)^2-1\right\}
\right].
\end{eqnarray}
Choosing $a_1=4a_2=-2a$, we obtain one parameter family of    
functions, 
\begin{eqnarray}
\label{Eq:hfoura}
 h^{(4)}_a(w)\equiv h_{\{-2a,\,-a/2\}}(w)
=\log\left\{1+2a-\frac{a}{8}
\left(w-\frac{1}{w}\right)^4\right\}.
\end{eqnarray}
The parameter $a$ is larger than or equal to $-1/2$ due to the
hermiticity condition of the classical solution.
For $a=-1/2$, we find 
\begin{eqnarray}
 \exp h^{(4)}_{-1/2}(w)=\frac{1}{16}\left(w-\frac{1}{w}\right)^4,
\end{eqnarray}
which has fourth order zeros at $w=\pm 1$.

The function $\exp h^{(4)}_a(w)$ 
can be rewritten as
\begin{eqnarray}
 \exp h^{(4)}_a(w)
=\frac{1}{(1+x)^2(1+y)^2}(1-x w^2)(1-x w^{-2})(1-y w^2)(1-y w^{-2}),
\label{Eq:h-4}
\end{eqnarray}
where the parameters $x$ and $y$ are related to $a$ by  
\begin{eqnarray}
\label{Eq:axy}
 a=
-\frac{8xy}{(1+x)^2(1+y)^2},\hspace{1cm}
(x+y)\left(1+\frac{1}{xy}\right)=4.
\end{eqnarray}
Since the function (\ref{Eq:h-4}) is invariant under the transformations
$x\rightarrow 1/x$ and $y\rightarrow 1/y$, we can impose the condition
$\abs{x}\leq 1$ and $\abs{y}\leq 1$ on the parameters $x$ and $y$.  It
follows from eq.~(\ref{Eq:axy}) that $\abs{x}< 1$ and $\abs{y}< 1$ for
$a>-1/2$, and $x=y=1$ for $a=-1/2$. Using $x$ and $y$, we can expand the
function $h_a^{(4)}(w)$ in a Laurent series,
\begin{eqnarray}
\label{Eq:hexp}
 h_a^{(4)}(w)=-\log(1+x)^2(1+y)^2
-\sum_{n=1}^\infty\,\frac{1}{n}\,(x^n+y^n)(w^{2n}+w^{-2n}).
\end{eqnarray}

In the theory expanded around the solution constructed with $h_a^{(4)}(w)$, 
the new BRS charge is given by
\begin{eqnarray}
\label{Eq:simtransBRS}
 \Q'(a)=Q(e^{h_a^{(4)}})-C\Bigl((\partial h_a^{(4)})^2e^{h_a^{(4)}}\Bigr).
\end{eqnarray}
We now examine if this charge $\Q'(a)$ is related to the original $\Q$ via 
a similarity transformation. Defining the operator
\begin{eqnarray}
 q(f)=\oint \frac{dw}{2\pi i}\, f(w)\,J_{\rm gh}(w),
\end{eqnarray}
and using the commutation relations \cite{rf:TT}
\begin{eqnarray}
[ q(f),~Q(g) ] &=& Q(fg)-2C(\partial f \partial g), \nonumber\\
~[ q(f),~C(g) ] &=& C(fg),
\label{comrel}
\end{eqnarray}
formally we obtain
\begin{eqnarray}
Q(e^{h_a^{(4)}})-C\Bigl((\partial h_a^{(4)})^2e^{h_a^{(4)}}\Bigr)= e^{q(h_a^{(4)})}\,\Q\, e^{-q(h_a^{(4)})}.
\label{simtransBRS1}
\end{eqnarray}
This equality holds of course 
only if the operator $\exp q(h_a^{(4)})$ is well-defined. 
Using (\ref{Eq:hexp}) and the mode expansion 
$J_{\rm gh}(w)=\sum q_n w^{-n-1}$, 
we expand the operator $q(h_a^{(4)})$ as
\begin{eqnarray}
 q(h_a^{(4)})=-\log(1+x)^2(1+y)^2\,q_0
-\sum_{n=1}^\infty \frac{1}{n}\,(x^n+y^n)(q_{2n}+q_{-2n}).
\end{eqnarray}

We find that the normal ordered form of $\exp q(h_a^{(4)})$ has a singularity
at $a=-1/2$, because the commutation relation between positive and
negative modes of $q(h_a^{(4)})$ diverges at $a=-1/2$ as follows,
\begin{eqnarray}
 \left[q^{(+)}(h_a^{(4)}),\,q^{(-)}(h_a^{(4)})\right]
=2\sum_{n=1}^\infty \frac{1}{n}(x^n+y^n)^2
\rightarrow 8\sum_{n=1}^\infty \frac{1}{n}=\infty\ \ \ 
(a\rightarrow -1/2).
\end{eqnarray}
Thus, (\ref{simtransBRS1}) holds for $a>-1/2$, but not for $a=-1/2$: the
expanded theory for $a>-1/2$ can be transformed back to the original
theory by the string field redefinition $\Psi'=\exp
q(h_a^{(4)})\times\Psi$, but this is not the case for $a=-1/2$. The
conclusion here is parallel to that for the case of the
function~(\ref{function-h}) \cite{rf:TT,rf:KT}.

It follows from the above argument that the function $h_{-1/2}^{(4)}(w)$
generates a new non-trivial solution with fourth order zeros. The
solution has a well-defined universal Fock space expression as for
the case of $h_a^m(w)$ \cite{rf:TT,rf:KT}.  Actually, substituting the
expressions
\begin{eqnarray}
&{ }&
 e^{h_{-1/2}^{(4)}(w)}-1=\frac{1}{16}(w^4+w^{-4})
-\frac{1}{4}(w^2+w^{-2})-\frac{5}{8},\nonumber\\
&{ }&\Bigl(\partial h_{-1/2}^{(4)}(w)\Bigr)^2
 e^{h_{-1/2}^{(4)}(w)}=
\frac{(1+w^2)^2(1-w^2)^2}{w^6}
\end{eqnarray}
into (\ref{Eq:sol}) and performing the integrations, we
obtain
\begin{eqnarray}
 \ket{\Psi_0(-1/2)}
&=&\frac{1}{8\pi}\sum_{n=0}^\infty (-1)^n
\left[\frac{1}{2n+5}+\frac{4}{2n+3}-\frac{10}{2n+1}
+\frac{4}{2n-1}+\frac{1}{2n-3}\right]Q_{-2n-1}\ket{I}\nonumber\\
&&+\frac{8}{3\pi}c_1\ket{I}+\frac{8}{5\pi}c_{-1}\ket{I}\nonumber\\
&&
-\frac{2}{\pi}\sum_{n=1}^\infty
(-1)^n\left[\frac{1}{2n+5}+\frac{1}{2n-3}
-\frac{2}{2n+1}\right]c_{-2n-1}\ket{I},
\label{new-sol}
\end{eqnarray}
where the mode expansions are  given by 
$J_{\rm B}(w)=\sum Q_n w^{-n-1}$ and
$c(w)=\sum c_n w^{-n+1}$. 

In the following, we will study the classical solutions obtained with
$h_a^{(4)}(w)$ in detail to provide physical interpretation of the
solution with higher order zeros.  We first discuss the cohomology of
the new BRS charge (\ref{Eq:simtransBRS}), and then the vacuum energy of
a non-perturbative vacuum.

\subsection{Cohomology}

For $a>-1/2$, the similarity transformation (\ref{Eq:simtransBRS})
relates the new BRS to the original BRS charge.  Then, the cohomology of
the new BRS charge is given by the similarity transformation of the
original cohomology \cite{rf:KO,rf:Henneaux}: the state $\ket{\psi}$
satisfying $\Q'(a)\ket{\psi}=0$ $(a>-1/2)$ can be written as
\begin{eqnarray}
 \ket{\psi}=e^{q(h_a^{(4)})}
\left(\ket{P}\otimes c_1\ket{0}
+\ket{P'}\otimes c_0c_1\ket{0}\right)+\Q'(a)\ket{\phi},
\end{eqnarray}
where $\ket{P}$ and $\ket{P'}$ are positive norm states in the matter
sector. Thus, there exists one-to-one correspondence between the
cohomologies of the new and original BRS charges for $a>-1/2$.

For $a=-1/2$, the new BRS charge is given by
\begin{eqnarray}
 \Q'(-1/2)&=&Q\left(\frac{1}{16}\left(
w-\frac{1}{w}\right)^4\right)-C\left(
w^{-2}\left(w^2-\frac{1}{w^2}\right)^2\right)
\nonumber\\
&=&\frac{3}{8}\Q
-\frac{1}{4}(Q_2+Q_{-2})+\frac{1}{16}\left(Q_4+Q_{-4}\right)
+2c_0-c_4-c_{-4}.
\end{eqnarray}
Using the first commutation relation in (\ref{comrel}), we can transform
the new BRS charge to a level four operator as
\begin{eqnarray}
\label{Eq:simtransQB}
e^{q(f)}\,\Q'\,e^{-q(f)}
=\frac{1}{16}(Q_4-16c_4),
\end{eqnarray}
where $f(w)$ and $q(f)$ are given by
\begin{eqnarray}
 f(w)=-4\log(1-w^{-2}),\ \ \ 
q(f)=4\sum_{n=1}^\infty\,\frac{1}{n}\,q_{-2n}.
\end{eqnarray}
Using eq.~(\ref{Eq:simtransQB}), we may find the cohomology for the case of
$a=-1/2$ in the same way as that for $h_a^m(w)$ \cite{rf:KT}.  First, the
cohomology of the operator $Q_4-16c_4$ becomes
\begin{eqnarray}
 \ket{P}\otimes b_{-4}b_{-3}b_{-2}\ket{0}
+\ket{P'}\otimes b_{-3}b_{-2}\ket{0}
+(Q_4-16c_4)\ket{\phi}.
\end{eqnarray}
This is due to the fact that $Q_4-16c_4$ may be obtained from the
original BRS charge $\Q$, by simply replacing the ghost oscillators $c_n$
and $b_n$ by $c_{n+4}$ and $b_{n-4}$ without changing their orders
\cite{rf:KT}.  Therefore, the state satisfying $\Q'(-1/2)\ket{\psi}=0$
can be written as
\begin{eqnarray}
 \ket{\psi}=
 \ket{P}\otimes e^{-q(f)}\,b_{-4}b_{-3}b_{-2}\ket{0}
+\ket{P'}\otimes e^{-q(f)}\,b_{-3}b_{-2}\ket{0}
+\Q'(-1/2)\ket{\phi}.
\end{eqnarray}
The ghost numbers of the cohomologically non-trivial states are $-3$ and
$-2$, which differ from that of the original BRS charge. Hence, we
conclude that there are no open string excitations perturbatively in the
theory expanded around the non-trivial solution with fourth order zeros.

\subsection{Vacuum energy}

The classical solution with $h_{-1/2}^{(4)}(w)$ is a non-trivial
solution as indicated by the cohomological analysis of the new BRS
charge.  Accordingly, the vacuum energy of the solution is expected to
have a non-zero value.  A solution with $a>-1/2$ is a pure gauge
solution and its vacuum energy must be zero.  

This resembles to the situation encountered in Ref.~\cite{rf:tomo}.  So,
it would be useful to recall what had happened in earlier case.  There,
due to some technical reasons, it was found difficult to directly
evaluate the energy density of our solution.  So rather than calculating
it directly, the theory expanded around a solution was taken and the
energy of a non-perturbative vacuum of that theory was evaluated.  The
action of the expanded theory is given by
\begin{eqnarray}
 S[\Phi]=-\frac{1}{g^2}\int
\left(\frac{1}{2}\Phi*\Q'(a)\Phi
+\frac{1}{3}\Phi*\Phi*\Phi\right),
\label{action for fluctuation}
\end{eqnarray}
where the new BRS charge is given as eq.(\ref{Eq:simtransBRS}).  The
action (\ref{action for fluctuation}) with a pure gauge solution is the
same as the original theory and it observes the tachyon vacuum as its
non-perturbative vacuum.  Therefore the vacuum energy takes a non-zero
value.  On the other hand, when expanded around the non-trivial vacuum,
the theory described by the action (\ref{action for fluctuation}) is
already at the tachyon vacuum and the vacuum energy is zero.  In
summary, by denoting the vacuum solution as $\Phi_0(a)$, the vacuum
energy density behaved as
\begin{eqnarray}
\label{Eq:expectvac}
f_a(\Phi_0(a))\equiv -S[\Phi_0(a)]/\,T_D\,V_D =
\left\{
\begin{array}{rl}
 -1 & (a>-1/2)\\
  0 & (a=-1/2),
\end{array}\right.
\label{vacuum energy}
\end{eqnarray}
where $T_D$ and $V_D$ are the D-brane tension and the space-time
volume, respectively.  In calculating $f_a(\Phi_0(a))$ in (\ref{vacuum
energy}), the level truncation approximation was utilized.  

We take the same approach to evaluate the vacuum energies for our new
solutions: we would like to see how the function $f_a(\Phi_0(a))$
behaves in the present case.  If it behaves differently from
eq.~(\ref{vacuum energy}), it implies the presence of a new vacuum.

In order to apply the level truncation analysis, let us calculate the
kinetic operator in the Siegel gauge:
\begin{eqnarray}
\label{Eq:kinopfour}
 L^{(4)}(a) 
   &=& {\cal T}(w e^{h_a^{(4)}})-q(w\,\partial e^{h_a^{(4)}})
-k\Bigl(w\,(\partial h_a^{(4)})^2\,e^{h_a^{(4)}}\Bigr),
\label{Eq:kinetic op}
\end{eqnarray}
where use was made of notations,
\begin{eqnarray}
 {\cal T}(f)=\oint \frac{dw}{2\pi i}f(w)T(w),\ \ \ 
 k(f)=\oint \frac{dw}{2\pi i}f(w).
\end{eqnarray}
The first two terms of (\ref{Eq:kinetic op}) is easily calculated as
\begin{eqnarray}
  {\cal T}(w e^{h_a^{(4)}})-q(w\,\partial e^{h_a^{(4)}})
&=& \left(1+\frac{5a}{4}\right)L_0
+\frac{a}{2}(L'_2+L'_{-2})-\frac{a}{8}(L'_4+L'_{-4}).
\end{eqnarray}
The primed operators $L_n'$ denote the twisted ghost Virasoro operators:
$L_n'\equiv L_n+n\,q_n+\delta_{n,0}$.
The Laurent expansion (\ref{Eq:hexp}) leads to
\begin{eqnarray}
 \partial h_a^{(4)}(w)=-4i w^{-1} \sum_{n=1}^\infty
 (x^n+y^n)\sin(2n\sigma),
\end{eqnarray}
where $w=e^{i\sigma}$. So, it follows that 
\begin{eqnarray}
 \Bigl(\partial h_a^{(4)}(w)\Bigr)^2e^{h_a^{(4)}(w)} &=&
 \partial h_a^{(4)}(w)\,\partial e^{h_a^{(4)}(w)}\nonumber\\
&=&
 32a w^{-2} \sum_{n=1}^\infty (x^n+y^n)\sin(2n\sigma)\sin^3\sigma
\cos\sigma.
\end{eqnarray}
Substituting this into (\ref{Eq:kinopfour}),
we calculate the last term as
\begin{eqnarray}
 -k\Bigl(w\,(\partial h_a^{(4)})^2\,e^{h_a^{(4)}}\Bigr)
=2a\left\{2(x+y)-(x^2+y^2)\right\}.
\end{eqnarray}

Now, we parameterize $x$ and $y$ as
\begin{eqnarray}
 x=\frac{t(1-t)}{1+t},\ \ y=\frac{t(1+t)}{1-t}.
\end{eqnarray}
Then, using $\xi=t^2$, we obtain the final expression for the
kinetic operator:
\begin{eqnarray}
\label{Eq:kinopfinal}
 L^{(4)}(a)
&=& \left(1+\frac{5a(\xi)}{4}\right)L_0
+\frac{a(\xi)}{2}(L'_2+L'_{-2})-\frac{a(\xi)}{8}(L'_4+L'_{-4})
+\alpha(\xi),
\end{eqnarray}
where $a(\xi)$ and $\alpha(\xi)$ are given by
\begin{eqnarray}
 a(\xi)&=& 
\frac{8\xi(1-\xi)^2}{(1-6\xi+\xi^2)^2},\nonumber\\
\alpha(\xi)&=&
\frac{32\xi^2(5+2\xi+\xi^2)}{(1-6\xi+\xi^2)^2}.
\end{eqnarray}
As $\xi$ changes from $-1$ to $3-2\sqrt{2}$, $a(\xi)$ varies from $-1/2$
to $+\infty$.

\begin{figure}[h]
\centerline{\includegraphics[width=12cm]{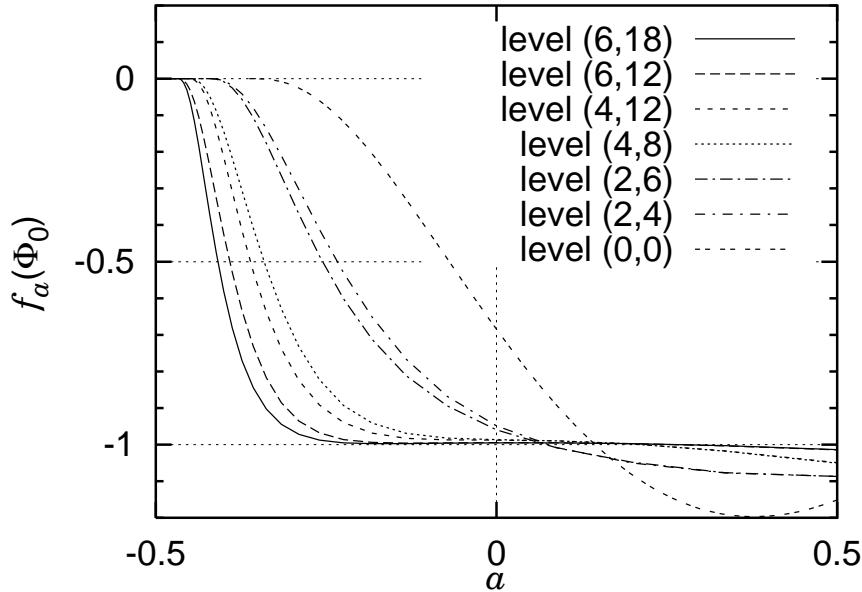}} \caption{Vacuum
energy.}  \label{fig:vacenergy}
\end{figure}

Under the Siegel gauge condition, the energy density is given by 
\cite{rf:tomo}
\begin{eqnarray}
 f_a(\Phi)
=2\pi^2\left(\frac{1}{2}
\langle \Phi,\,c_0L^{(4)}(a)\Phi\rangle
+\frac{1}{3}\langle \Phi,\,\Phi*\Phi\rangle \right).
\end{eqnarray}
At level $(0,0)$ truncation, the component field is $t\,c_1\ket{0}$, and 
the energy density is
\begin{eqnarray}
 f_a(t)&=&2\pi^2\left(-\frac{1}{2}\,\lambda(\xi)\,t^2
+\frac{1}{3}\,\left(\frac{3\sqrt{3}}{4}\right)^3 t^3\right),\nonumber\\
\lambda(\xi)&=&
1+\frac{5a(\xi)}{4}-\alpha(\xi).
\end{eqnarray}
$\lambda(\xi)$ has two real roots 
$\xi^+=0.0759112\cdots$ and $\xi^-=-0.0933769\cdots$,
and these correspond to $a(\xi^+)=1.89932\cdots$ and
$a(\xi^-)=-0.362772\cdots$, respectively.
We find that $\lambda(\xi)>0$ if $\xi^-<\xi<\xi^+$ and
$\lambda(\xi)<0$ if $\xi<\xi^-$ or $\xi>\xi^+$. Then, the energy density
has a local minimum as follows
\begin{eqnarray}
 f_a(t_0)=\left\{
\renewcommand{\arraystretch}{1.5}
\begin{array}{ll}
\displaystyle
 -\frac{\pi^2}{3}\left(\frac{4}{3\sqrt{3}}\right)^6\lambda(\xi)& 
  (\xi^-\leq \xi\leq \xi^+)\\
 0 & (-1\leq \xi <\xi^-\ \ {\rm or}\ \ 
      \xi^+<\xi<3-2\sqrt{2}).
\end{array}\right.
\renewcommand{\arraystretch}{1}
\end{eqnarray}
We proceeded up to the level $(6,18)$ and evaluated the local minimum of
the energy density. The resulting vacuum energy is depicted in
Fig.~\ref{fig:vacenergy}. We observe that the vacuum energy approaches
to the function (\ref{Eq:expectvac}) as the truncation
level is increased.  

The result in this section, together with the cohomology analysis, is
consistent with the statement that the non-trivial solution with the
forth order zeros describes the same tachyon vacuum as other non-trivial
solutions obtained earlier.

\section{Summary and outlook}

In the construction of classical solutions to CSFT, each solution is
specified by an underlying holomorphic function $F(w) = e^{h(w)}$.
Because of the inversion relation $F(w)=F(-1/w)$, the same numbers of
zeros are distributed inside and outside the unit circle $|w|=1$.  The
solutions with functions whose zeros are located off the unit circle are
pure gauge, while the non-trivial solutions, that cannot be gauged away,
are obtained with functions with zeros all on the unit circle $|w|=1$.
The functions $F_{-1/2}^{m}(w)= \exp \Bigl(h_{-1/2}^{m}(w)\Bigr)$ where
$h_{a}^{m}(w)$ are given in (\ref{function-h}) have second-order zeros.
In \cite{rf:Drkr2}, a conjecture has been made that the order of zeros
is related to the number of D-branes and therefore different orders of
zeros will correspond to different vacua of the theory.

In this letter, we have constructed a non-trivial solution specified by
a function with the fourth order zeros. The cohomological argument on the
new BRS charge shows the absence of open string excitations at the
perturbative level. Our numerical analysis on the vacuum energy
indicates that the non-trivial solution with the fourth order zeros
describes the tachyon condensed vacuum.  From these results we conclude
that our new non-trivial solutions correspond to the same tachyon vacuum
described by the solutions reported earlier \cite{rf:TT,rf:KT} and they are counter examples to the above conjecture.

We have reported yet another analytic solutions to the
CSFT.  Our results suggest that all these classical solutions as well as
those obtained earlier are sound candidates describing the same tachyon
vacuum.  This is consistent only if these solutions are gauge
equivalent.  We will report how they could be related each other in the
forthcoming paper \cite{rf:igarashi}.

\section*{Acknowledgements}

This work is supported in part by the Grants-in-Aid for Scientific
Research No. 13135209, 15540262 from the Japan Society for the
Promotion of Science.  The authors thank the Yukawa Institute for
Theoretical Physics at Kyoto University. Discussions during the YITP
workshop YITP-W-04-03 on ``Quantum Field Theory 2004" 
were useful to complete this work.  


\end{document}